%% file: Jevicki4.tex
\documentclass[12pt]{article}
\usepackage{epsfig}
\usepackage{amsmath}
\usepackage{amsfonts}
\usepackage{amssymb}
\usepackage{graphicx}

\input econfmacros.tex
\def\Title#1{\begin{center} {\Large {\bf #1} } \end{center}}

\begin{document}

\Title{Bi-local Model of AdS/CFT \\ and Higher Spin Gravity}

\bigskip\bigskip


\begin{raggedright}  

{\it Antal Jevicki \footnote{Based on talks given at the ``Eleventh workshop on non-perturbative Quantum Chromodynamics'', l'Institut Astrophysique de Paris, France, June 6-10, 2011; the 35$^{\rm th}$ Johns Hopkins workshop on AdS/CFT and its applications, Budapest, Hungary, June 22-24, 2011.}, Kewang Jin \footnote{Present address: Institut f\"{u}r Theoretische Physik, ETH-Z\"{u}rich, CH-8093 Zurich, Switzerland.}, Qibin Ye \\
Department of Physics\\
Brown University, Box 1843\\
Providence, RI 02912, USA}
\bigskip\bigskip
\end{raggedright}


\section{Introduction}

The quantum field theory of the $O(N)$ vector model
\begin{equation}
L=\int d^d x {1 \over 2}(\partial_\mu \phi^i)(\partial^\mu \phi^i)+V(\phi \cdot \phi), \qquad i=1,...,N
\end{equation}
at its fixed points represents a well-known field theory for critical phenomena in three dimensions. More recently this field theory was used as a particularly simple model of AdS$_{d+1}$/CFT$_d$ correspondence with higher spin theory. In three dimensions, besides the trivial free field UV fixed point, one also has a nontrivial IR fixed point. For both fixed points an AdS/CFT duality with Vasiliev's higher spin theory \cite{Vasiliev:1995dn, Vasiliev:2003ev} in four dimensions was pointed out \cite{Klebanov:2002ja, Sezgin:2002rt}. This was explicitly confirmed through a comparison of boundary correlators by Giombi and Yin \cite{Giombi:2009wh, Giombi:2010vg}. Furthermore a construction of the AdS spacetime and higher-spin fields from the CFT was outlined in \cite{Das:2003vw, Koch:2010cy}. In this presentation we review the elements of the constructive approach of \cite{Das:2003vw, Koch:2010cy} and provide some further comparisons. A related work \cite{Douglas:2010rc} employs a renormalization group equation but we will not use this scheme here. A full list of references is beyond the scope of the summary and can be found in the published articles.
 
The comparison of correlators involves conformal primary operators given by the classically conserved, traceless and symmetric currents
\begin{equation}
J_{\mu_1 \mu_2 \cdots \mu_s}=\sum_{k=0}^s (-1)^k \left( \begin{array}{c} s-1/2 \\ k \end{array} \right) \left( \begin{array}{c} s-1/2 \\ s-k \end{array} \right) \partial_{\mu_1} \cdots \partial_{\mu_k} \phi \, \partial_{\mu_{k+1}} \cdots \partial_{\mu_s} \phi-\text{traces}.
\end{equation} 
These operators can be summarized in the semi bi-local form
\begin{equation}
{\mathcal{O}}(x,\epsilon)=\phi^i(x-\epsilon)\sum_{n=0}^\infty \frac{1}{(2n)!}(2\epsilon^2 \overleftarrow{\partial_x} \cdot \overrightarrow{\partial_x}-4(\epsilon \cdot \overleftarrow{\partial_x})(\epsilon \cdot \overrightarrow{\partial_x}))^n \phi^i(x+\epsilon).
\end{equation}
The direct approach of reconstructing the AdS theory \cite{Das:2003vw, Koch:2010cy} is based on the notion of bi-local collective fields
\begin{equation}
\Psi(x_1^\mu,x_2^\nu)=\sum^N_{i=1}\phi^i(x_1^\mu)\phi^i(x_2^\nu).
\end{equation}
They represent a most general set of $O(N)$ invariants and are suitable collective variables for a full description of the large $N$ theory. The collective fields $\Psi(x_1^\mu,x_2^\nu)$ stand as a more general set than the conformal fields $\mathcal{O}(x^\mu,\epsilon^\nu)$ due to the fact that one requires $\epsilon^2=0$ for the later, in accordance with the traceless condition. More importantly, the bi-local fields close a set of large $N$ Schwinger-Dyson equations, a fact which implies the existence of an effective action. The action and the associated bi-local field representation are known to allow for a systematic $1/N$ expansion. It is a proposal of \cite{Das:2003vw, Koch:2010cy} that this bi-local effective action provides a bulk description of the AdS$_4$ dual higher-spin gravity. In particular, for the three dimensional critical CFT a procedure for constructing the AdS phase space theory was given in the light-cone gauge framework \cite{Koch:2010cy}. It specified a one-to-one operator map expressing the bulk light-cone higher spin fields in terms of the bi-local fields through the transform
\begin{eqnarray}
\Phi(x^-,x,z,\theta)&=&\int dp^+ dp^x dp^z e^{i (x^- p^++x p^x+z p^z)} \cr
&&\int dp_1^+ dp_2^+ dp_1 dp_2 \delta(p_1^+ + p_2^+ - p^+)\delta(p_1+p_2-p^x) \cr
&&\delta\Bigl(p_1 \sqrt{p_2^+ / p_1^+} -p_2 \sqrt{p_1^+ / p_2^+}-p^z\Bigr) \cr
&&\delta\bigl(2\arctan\sqrt{p_2^+ / p_1^+}-\theta\bigr) \tilde{\Psi}(p_1^+,p_2^+,p_1,p_2),
\label{LCmapping4}
\end{eqnarray}
where $\tilde{\Psi}(p_1^+,p_2^+,p_1,p_2)$ is the Fourier transform of the bi-local field $\Psi(x_1^-,x_2^-,x_1,x_2)$.

The physical mechanism behind the explicit correspondence is then seen to be given by a bi-particle system of the collective dipole which through a canonical transformation maps into the first quantized version of the higher spin system. Details of the Dipole/AdS correspondence were given in \cite{Jevicki:2011ss} where a covariant description was also discussed. It can be shown that a $2d$ dimensional phase space of the ``collective dipole'' maps identically into to the phase space of the AdS$_{d+1}$ particle with continuous spins. The bi-particle dipole therefore provides a ``world-sheet'' description of the AdS/CFT correspondence in the higher-spin case.

In what follows we will first give a short review of this bi-local mechanism. We will then in the second part of the lecture present some results concerning the full non-linear higher spin theory. One of the basic puzzles regarding the correspondence involves the fact that the bi-local theory represents a symmetrical $2d$ dimensional description of the theory, on the other hand the full gauge invariant higher spin theory of Vasiliev is represented in $d+1$ dimensional AdS spacetime with further extra coordinates generating the higher-spin fields. We demonstrate that there exists a gauge fixing procedure that brings Vasiliev's theory to a symmetrical form with the counting of degrees of freedom agreeing with that of the bi-local theory. This implies that the ultimate full agreement at the nonlinear level can be contemplated through a standard change of field variables. We will mention some implications of this fact on the properties of the higher spin theory in the conclusion.

\section{From Collective to Higher Spin Fields}

\begin{figure}
\begin{center}
\includegraphics[width=.75\textwidth]{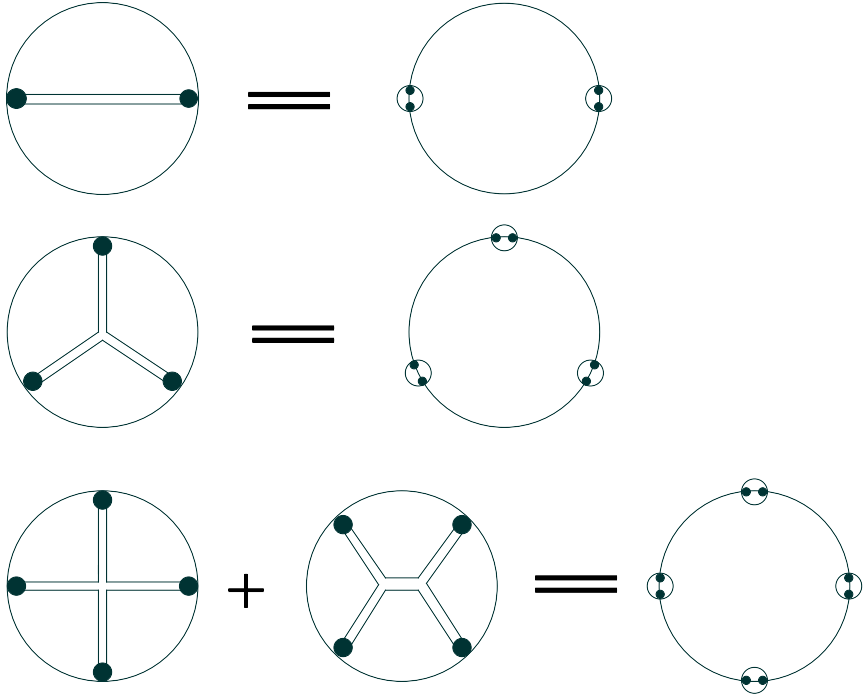}
\caption{Illustration of two-, three-, four-point collective field diagrams.}
\label{diagram}
\end{center}
\end{figure}

The basis of the direct approach \cite{Das:2003vw} to construct the dual AdS theory comes from the fact that one is able to write the partition function entirely in terms of the bi-local fields $\Psi(x^\mu,y^\nu)$, namely
\begin{equation}
Z=\int [d\phi^i(x)] e^{-S[\phi]}=\int \prod_{x,y} d\Psi(x,y) \mu(\Psi) e^{-S_c[\Psi]}=\int [d\Psi] e^{-S_{\rm eff}}
\end{equation}
where $\mu(\Psi)$ is a computable measure and the collective (effective) action is given by
\begin{equation}
S_{\rm eff}=\int d^d x [-\frac{1}{2}\partial_x^2 \Psi(x,y)\vert_{x=y}+\frac{\lambda}{4} \Psi^2(x,y)\vert_{x=y}] -\frac{N}{2} {\rm Tr} \ln \Psi.
\end{equation}
Here the trace is defined as ${\rm Tr} \, B=\int d^dx B(x,x)$ and the bi-local construction introduces the star product of the form
\begin{equation}
(\Psi \star \Phi)(x,y)=\int dz \Psi(x,z) \Phi(z,y).
\end{equation}
The origin of the $\ln \Psi$ interaction in the effective action (and also of the measure factor in the integral) arises from the Jacobian
\begin{equation}
\int [d\phi^i(x)] e^{-S} \rightarrow \int [d\Psi] \det \left\vert \frac{\partial \phi^i(x)}{\partial \Psi(x,y)} \right\vert e^{-S}
\end{equation}
which is computable \cite{Das:2003vw}. Altogether the effective action can be seen of order $\mathcal{O}(N)$ while the measure is $\mathcal{O}(1)$. This implies that in the bi-local representation, the role of coupling constant is played by $1/N$. To generate the expansion one first solves the equation of motion provided by $S_{eff}$
\begin{equation}
\Psi \frac{\partial S_{\rm eff}}{\partial \Psi}=0 \Rightarrow -\partial_x^2\Psi_0(x,y)+\lambda \Psi_0^2(x,y)-N\delta(x-y)=0.
\end{equation}
The $1/N$ perturbation theory follows by expanding around the saddle point
\begin{equation}
\Psi=\Psi_0+\frac{1}{\sqrt{N}} \eta.
\end{equation}
The quadratic term in $\eta$ defines the propagator of the bi-local field
\begin{equation}
S_2[\eta]=\frac{\lambda}{4 N}\int d^d x \, \eta(x,x)\eta(x,x)+\frac{1}{4}{\rm Tr}[\Psi_0^{-1} \, \eta \, \Psi_0^{-1} \, \eta]
\end{equation}
while higher order terms define the vertices (for more detailed studies and references to collective field theory, see \cite{Das:2003vw}). The manner in which the correlation functions of the $O(N)$ model are recovered in the bi-local theory is strikingly reminiscent of the Witten type diagrams in AdS space. The diagrams in Figure \ref{diagram} emphasize this point.

In summary, this represents a highly non-trivial rearrangement of perturbation theory and offers an intuitive basis for the proposal that the bi-local representation is capable of giving a bulk description of the AdS theory. This expectation is made explicit by establishing a one-to-one correspondence to the bulk fields. To describe this result \cite{Jevicki:2011ss}, we need some basic features of higher spin fields in AdS \cite{Flato:1978qz, Fronsdal:1978vb}.

A spin $s$ field is represented by a symmetric and double traceless tensor of rank $s$: $h_{\mu_1...\mu_s}(x^\mu)$, which obeys the equations of motion
\begin{eqnarray}
\nabla_\rho \nabla^\rho h_{\mu_1...\mu_s}-s\nabla_\rho \nabla_{\mu_1}h^\rho_{~\mu_2...\mu_s}+\frac{1}{2}s(s-1)\nabla_{\mu_1} \nabla_{\mu_2}h^\rho_{~\rho \mu_3...\mu_s} \cr
+2(s-1)(s+d-2)h_{\mu_1...\mu_s}=0.
\label{HSeom}
\end{eqnarray}
The gauge transformation is given by
\begin{equation}
\delta_{\Lambda}h^{\mu_1...\mu_s}=\nabla^{\mu_1}\Lambda^{\mu_2...\mu_s}, \qquad g_{\mu_2 \mu_3} \Lambda^{\mu_2...\mu_s}=0.
\end{equation}
A covariant gauge can be specified with the gauge conditions
\begin{equation}
\nabla^\rho h_{\rho \mu_2...\mu_s}=0, \qquad g^{\rho \sigma}h_{\rho \sigma \mu_3...\mu_s}=0.
\end{equation}
Then the equation of motion (\ref{HSeom}) reduces to
\begin{equation}
(\square+m^2)h_{\mu_1...\mu_s}=0,
\end{equation}
with the effective mass $m^2=s^2+(d-5)s-2(d-2)$.

It is useful to embed the $d+1$ dimensional AdS spacetime $x^\mu$ into $d+2$ dimensional hyperboloid $x^\alpha$. The higher spin field $h_{\mu_1...\mu_s}(x^\mu)$ is related to the after-embedding higher spin field $k_{\alpha_1...\alpha_s}(x^\alpha)$ by
\begin{equation}
k_{\alpha_1 ... \alpha_s}(x^\alpha)=x^{~\mu_1}_{\alpha_1} \cdots x^{~\mu_s}_{\alpha_s} h_{\mu_1 ... \mu_s}(x^\mu),
\end{equation}
where $x^{~\mu}_{\alpha} \equiv \partial x^\mu / \partial x^\alpha$. Introducing an internal set of coordinates $y^\alpha$ spacetime, one forms the field with all spins
\begin{equation}
K(x^\alpha,y^\alpha) \equiv \sum_s k_{\alpha_1...\alpha_s}(x^\alpha)y^{\alpha_1} \cdots y^{\alpha_s}.
\end{equation}
In this notation the constraints implied by embedding, the covariant gauge conditions as well as the equations of motion become the following system of equations
\begin{eqnarray}
&&\partial_x^2 K(x,y)=0, \label{froncons1} \\
&&\partial_y^2 K(x,y)=0, \\
&&\partial_x \cdot \partial_y K(x,y)=0, \\
&&(x \cdot \partial_x+y \cdot \partial_y+1)K(x,y)=0, \\
&&x \cdot \partial_y K(x,y)=0. \label{froncons5}
\end{eqnarray}
It is easy to check that the constraints (\ref{froncons1}-\ref{froncons5}) are all first-class constraints. From here one can find the light-cone higher spin field $\Phi(x^-,x,z,\theta)$ in (\ref{LCmapping4}) through further gauge fixing of $K(x^\alpha,y^\alpha)$ in AdS$_4$.

Continuing with the covariant gauge representation, we see a clear asymmetry between the spacetime coordinates $x$ and the internal spin coordinates $y$. This is in contrast with the bi-local field representation of the $O(N)$ CFT. A relevant nontrivial fact is that one can through a series of canonical transformations achieve a totally symmetric description
\begin{equation}
\Phi(p,q)=(FK)(x,y)
\end{equation}
where $p=(x+y)/2$, $q=(x-y)/2$ and the kernel for a particular spin $s$ is given by
\begin{equation}
F_s=\sum_k (4^k k!)^{-1} (y \cdot \partial_x)^{2k}/(\hat{n}+1)(\hat{n}+2)\cdots(\hat{n}+k)
\label{Fronsmap}
\end{equation}
with $\hat{n}=y \cdot \partial_y$. After the mapping (\ref{Fronsmap}), as well as a Fourier transformation
\begin{equation}
\Phi(u,v)=\int dp dq \, e^{i p \cdot u+i q \cdot v} \Phi(p,q),
\end{equation}
one finds the symmetric version 
\begin{eqnarray} 
&&(u \cdot \partial_u+1/2)\Phi(u,v)=0, \label{fronsym1} \\
&&(v \cdot \partial_v+1/2)\Phi(u,v)=0, \label{fronsym2} \\
&&u^2=0, \label{fronsym3} \\
&&v^2=0, \label{fronsym4} \\
&&u \cdot v=0. \label{fronsym5}
\end{eqnarray} 

Now we have a symmetrical description of the linearized higher spin theory which is now suitable for establishing a one-to-one map to the bi-local fields. To do that one still has to reduce the system by solving the constraints. The first four constraints (\ref{fronsym1}-\ref{fronsym4}) which separately involve the coordinates $u$ and $v$ are solved as follows. One parameterizes the cone $u^2=0$ as
\begin{equation}
u_0=U\sin t, \qquad u_{d+1}=U\cos t, \qquad \vec{u}=U\hat{u},
\end{equation}
with $\hat{u}^2=1$, we find that the constraint (\ref{fronsym1}) becomes $\partial / \partial U+1/2=0$. Consequently the dependence on the variable $U$ can be factored out
\begin{equation}
\phi(u)=U^{-1/2}\phi(t,\hat{u}),
\end{equation}
and the remaining degrees of freedom are the coordinates $(t,\hat{u})$ (and their conjugates). Similarly, this reduction works for the $v$ system. Therefore, by solving the first four constraints, we reduced the $\Phi(u,v)$ with $2(d+2)$ coordinates to the unconstrained space featuring $2d$ variables. This agrees precisely with the bi-local collective field $\Psi(x^\mu,y^\nu)$ in $d$ dimensions. 

To agree with the form of the collective field equations of \cite{Das:2003vw} one next considers the one remaining constraint of Fronsdal and replaces it by the form
\begin{equation}
\partial_u^2 \partial_v^2 \Phi(u,v)=0.
\end{equation}
The two forms can be seen not to commute with each other, consequently this can be seen as a change of gauge conditions in the higher spin theory, with the result that equation of motion now agrees with the collective field equation. This shows that the bi-local collective field equation corresponds in a linearized approximation to a specific gauge fixing of Fronsdal's covariant  higher spin theory.

\section{A Symmetric Gauge of Higher Spin Theory}

In the above we have described the correspondence between the bi-local large $N$ field theory and a linearized higher spin theory in one higher dimensional AdS space \cite{Jevicki:2011ss}. The one-to-one map was established through a canonical transformation involving the phase space of the collective dipole and the phase space of the higher spin particle in AdS. From the viewpoint of higher spin theory it was the ability to reformulate the theory in a $2d$ symmetrical coordinate space. This was done by transforming Fronsdal's linearized higher spin field equations into a symmetric form.

As described earlier the bi-local field theory deduced from the vector model is fully known at the nonlinear level, the nonlinearity being governed by $1/N$ as its coupling constant. On the AdS side, the remarkable construction of Vasiliev represents a closed set of nonlinear higher spin equations. But one has a huge discrepancy concerning the coordinate spaces of the two theories. For example in the case of AdS$_4$, Vasiliev's equations are expressed in a $4+8$ dimensional space where the extra eight coordinates parametrize the sequence of higher spin fields with an exact higher-spin gauge symmetry. For a potential exact correspondence with the $3+3$ dimensional bi-local field theory one would like to demonstrate that through gauge fixing one can reduce Vasiliev's theory to a symmetrical $3+3$ dimensional base space involving a single scalar field. We will now show the existence of such a gauge and present a reduced scalar field representation of Vasiliev's theory. The steps are in part analogous to a similar gauge fixing/reduction known in self-dual Yang-Mills theory.

Vasiliev's higher-spin equations in AdS$_4$ are given by \cite{Vasiliev:1995dn}
\begin{eqnarray}
&&dW=W*W, \label{v3eom1} \\
&&dB=W*B-B*\pi(W), \label{v3eom2} \\
&&dS=W*S-S*W, \label{v3eom3} \\
&&S*B=B*\pi(S), \label{v3eom4} \\
&&S*S=dz^\alpha dz_\alpha(i+B*\kappa)+d\bar{z}^{\dot{\alpha}} d\bar{z}_{\dot{\alpha}}(i+B*\bar{\kappa}), \label{v3eom5}
\end{eqnarray}
where $\kappa=\exp(i z_\alpha y^\alpha)$, $\bar{\kappa}=\exp(i \bar{z}_{\dot{\alpha}} \bar{y}^{\dot{\alpha}})$ and the symmetry operator $\pi$ changes the signs of all undotted spinors
\begin{equation}
\pi(f)(dz,d\bar{z};z,\bar{z};y,\bar{y})=f(-dz,d\bar{z};-z,\bar{z};-y,\bar{y}).
\end{equation}
Similarly, one can define the operator $\bar{\pi}$ which changes the signs of all dotted spinors. The star product law is defined as
\begin{equation}
(f*g)(Z;Y)=\int d^4 U d^4 V \exp[i(u_\alpha v^\alpha+\bar{u}_{\dot{\alpha}} \bar{v}^{\dot{\alpha}})] f(Z+U;Y+U) g(Z-V;Y+V),
\label{star3}
\end{equation}
where $U=(u_\alpha,\bar{u}_{\dot{\alpha}})$, $V=(v_\alpha,\bar{v}_{\dot{\alpha}})$ are the integration variables. This product law is some particular symbol version of the Heisenberg-Weyl algebra
\begin{equation}
[y_\alpha,y_\beta]_*=-[z_\alpha,z_\beta]_*=2i\epsilon_{\alpha\beta} \ , \qquad [\bar{y}_{\dot{\alpha}},\bar{y}_{\dot{\beta}}]_*=-[\bar{z}_{\dot{\alpha}},\bar{z}_{\dot{\beta}}]_*=2i\epsilon_{\dot{\alpha}\dot{\beta}} \ ,
\label{comm3}
\end{equation}
(all other commutators vanish). Furthermore, $\kappa$, $\bar{\kappa}$ possess the properties of the inner Kleinian operator
\begin{eqnarray}
&&\kappa*f(z,y)=f(-z,-y)*\kappa=\kappa f(y,z), \\
&&\bar{\kappa}*g(\bar{z},\bar{y})=g(-\bar{z},-\bar{y})*\bar{\kappa}=\bar{\kappa}g(\bar{y},\bar{z}), \\
&&\kappa*\kappa=1, \qquad \bar{\kappa}*\bar{\kappa}=1.
\end{eqnarray}
It is important that the nonlinear equations are explicitly invariant under the higher-spin gauge transformations
\begin{eqnarray}
&&\delta W=d\epsilon-W*\epsilon+\epsilon*W, \\
&&\delta S=\epsilon*S-S*\epsilon, \\
&&\delta B=\epsilon*B-B*\pi(\epsilon).
\end{eqnarray}
One can also check the hermitian conjugates (involutions) of the higher-spin fields are
\begin{equation}
W^\dagger=-W, \qquad S^\dagger=-S, \qquad B^\dagger=-\tilde{B},
\label{herm31}
\end{equation}
where we used the hermiticity properties of the variables
\begin{equation}
(z_\alpha)^\dagger=-\bar{z}_{\dot{\alpha}}, \quad (y_\alpha)^\dagger=\bar{y}_{\dot{\alpha}}, \quad (dz_\alpha)^\dagger=d\bar{z}_{\dot{\alpha}}, \quad (dx_\nu)^\dagger=dx_\nu, \quad \kappa^\dagger=\bar{\kappa}.
\end{equation}

Looking at equations (\ref{v3eom1}), $W$ is a flat connection in spacetime, at least locally we can solve for $W$ and fix a gauge to set $W$ to zero. We will denote by $S'$ and $B'$ the corresponding master fields in this gauge. The equations of motion (\ref{v3eom2}, \ref{v3eom3}) then states that $S'$ and $B'$ are independent of the spacetime coordinate $x^\mu$. Explicitly, we have the gauge transformation
\begin{eqnarray}
W(x \vert Y,Z)&=&g^{-1}(x \vert Y,Z)*d_x g(x \vert Y,Z), \\
S(x \vert Y,Z)&=&g^{-1}(x \vert Y,Z)*S'(Y,Z)*g(x \vert Y,Z), \\
B(x \vert Y,Z)&=&g^{-1}(x \vert Y,Z)*B'(Y,Z)*\pi(g)(x \vert Y,Z).
\end{eqnarray}
According to the hermiticity condition (\ref{herm31}), the gauge function must satisfy
\begin{equation}
g^\dagger=g^{-1}.
\end{equation}
The equations for $S'$ and $B'$ now take the form
\begin{eqnarray}
&&S'*S'=dz^\alpha dz_\alpha (i+B'*\kappa)+d\bar{z}^{\dot{\alpha}} d\bar{z}_{\dot{\alpha}}(i+B'*\bar{\kappa}), \\
&&S'*B'=B'*\pi(S').
\end{eqnarray}
Omitting the primes and after a shift of the $S$ field
\begin{equation}
S_\alpha \to z_\alpha+S_\alpha, \qquad \bar{S}_{\dot{\alpha}} \to \bar{z}_{\dot{\alpha}}+\bar{S}_{\dot{\alpha}},
\end{equation}
we find the equations of motion in components
\begin{eqnarray}
&&i\partial_\alpha S^\alpha-S_\alpha * S^\alpha=B*\kappa, \label{VSet51} \\
&&i\bar{\partial}_{\dot{\alpha}}\bar{S}^{\dot{\alpha}}-\bar{S}_{\dot{\alpha}}*\bar{S}^{\dot{\alpha}}=B*\bar{\kappa},
\label{VSet52} \\
&&i\partial_\alpha\bar{S}_{\dot{\beta}}-i\bar{\partial}_{\dot{\beta}}S_\alpha-[S_\alpha,\bar{S}_{\dot{\beta}}]_*=0,
\label{VSet53} \\
&&i\partial_\alpha B-S_\alpha*B-B*\pi(S_\alpha)=0, \label{VSet54} \\
&&i\bar{\partial}_{\dot{\alpha}}B-\bar{S}_{\dot{\alpha}}*B-B*\bar{\pi}(\bar{S}_{\dot{\alpha}})=0. \label{VSet55}
\end{eqnarray}

This description of Vasiliev's theory was already used by Giombi and Yin in their second evaluation of three-point correlators. The equations in question (known as the $W=0$ gauge) involve a scalar field $B$ and a vector $S$. Our goal is to gauge fix and reduce this set of equations even further and show a description in terms of a single scalar field. To this end we follow some steps known from analogous treatment of self-dual Yang-Mills theory.

We note first, that the last two equations (\ref{VSet54}, \ref{VSet55}) for the $B$ field are not independent. Using the first two equations (\ref{VSet51}, \ref{VSet52}), one can solve for the $B$ field and check that (\ref{VSet54}, \ref{VSet55}) are satisfied. Therefore, we can totally eliminate the $B$ field and find the following five equations for the $S$ field
\begin{eqnarray}
&&F_{\alpha\dot{\beta}} \equiv i\partial_\alpha\bar{S}_{\dot{\beta}}-i\bar{\partial}_{\dot{\beta}}S_\alpha-[S_\alpha,\bar{S}_{\dot{\beta}}]_*=0, \\
&&(i\partial_\alpha S^\alpha-S_\alpha * S^\alpha)*\kappa-(i\bar{\partial}_{\dot{\alpha}}\bar{S}^{\dot{\alpha}}-\bar{S}_{\dot{\alpha}}*\bar{S}^{\dot{\alpha}})*\bar{\kappa}=0. \label{reality}
\end{eqnarray}
The last equation (\ref{reality}) represents the reality condition for the $B$ field.

We next introduce an ansatz
\begin{eqnarray}
S_1=-i M^{-1}*\partial_1 M, &\quad& S_2=-i\bar{M}*\partial_2\bar{M}^{-1}, \\
\bar{S}_{\dot{1}}=-i\bar{M}*\bar{\partial}_{\dot{1}} \bar{M}^{-1}, &\quad& \bar{S}_{\dot{2}}=-i M^{-1}*\bar{\partial}_{\dot{2}} M,
\end{eqnarray}
where we used the notation
\begin{equation}
M^\dagger=\bar{M}.
\end{equation}
This ansatz solves the $F_{1\dot{2}}=F_{2\dot{1}}=0$ equations automatically. Next, by introducing an invariant scalar field $J=M*\bar{M}$, the other two equations $F_{1\dot{1}}=F_{2\dot{2}}=0$ take a simple form
\begin{eqnarray}
&&\bar{\partial}_{\dot{1}}(J^{-1}*\partial_1 J)=0, \label{Vzero1} \\
&&\partial_2(J^{-1}*\bar{\partial}_{\dot{2}} J)=0. \label{Vzero2}
\end{eqnarray}
Finally the last equation $F_{12}*\kappa-F_{\dot{1}\dot{2}}*\bar{\kappa}=0$ becomes
\begin{equation}
\partial_2(J^{-1}*\partial_1 J)*\kappa+\bar{\partial}_{\dot{1}}(J^{-1}*\bar{\partial}_{\dot{2}} J)*\bar{\kappa}=0,
\label{Vzero3}
\end{equation}
where we used the symmetry property $M(Y,Z)=M(-Y,-Z)$ in the bosonic case.

To summarize, we have through gauge fixing and elimination of fields reduced Vasiliev's theory to that of a single scalar field $J$. This field is hermitian
\begin{equation}
J^\dagger=J
\end{equation}
and it matches up with the properties of the bi-local collective field. Of the three equations deduced above (\ref{Vzero1}-\ref{Vzero3}), we interpret the first two as constraints, while the third would be the equation of motion. The constraint equations appear to reduce the base space of the scalar field $J$ from $4+4$ dimensions to $3+3$ dimensions in agreement with the bi-local dipole coordinate space.

We note that our reformulation in terms of a single scalar field $J$ giving a relatively simple description of Vasiliev's equations might also be useful for writing an action for the higher spin theory. This represents an interesting problem (for recent approach, see \cite{Boulanger:2011dd}).

\section{Conclusion}

We have given a short description of the construction of AdS/CFT correspondence involving the $O(N)$ vector theories and higher spin AdS gravity. This approach is based on the bi-local collective field representation which as we argue stands in a one-to-one relationship with the bulk fields of AdS. The construction was presented in the case of a scalar $O(N)$ vector theory but it applies in a similar way to the fermionic theory also. Very interesting would be the extension of this approach to dualities involving two dimensional conformal models which enjoy a similar correspondence \cite{arXiv:1011.2986, arXiv:1106.1897, arXiv:1008.4579, arXiv:1107.0290}.

The bi-local field representation possesses several relevant features which have implications on the nature of higher spin theory. First of all the bi-local theory faithfully reproduces all finite temperature properties of the $O(N)$ field theory, in particular the two phases identified recently in \cite{arXiv:1109.3519}. The high temperature phase was already exhibited in \cite{Das:2003vw} while the lower phase can be naturally seen in the Hamiltonian formulation of the theory. Next, in this approach the relationship between the CFT operators (bi-locals) and bulk AdS fields is ``just'' a nonlinear field transformation. This fact has some relevant implications on the properties of the nonlinear higher spin theory of Vasiliev. In the case of the free $O(N)$ vector model correspondence it can be shown that the seemingly nonlinear collective field theory can be linearized through the existence of a nonlinear field redefinition. This implies an identical property for the corresponding Vasiliev's theory, namely a possibility of its ``linearization''. Such feature for the particular higher spin theory (which is dual to the free CFT) has been put forward recently by Maldacena and Zhiboedov \cite{arXiv:1112.1016}. They demonstrate it as a consequence of the infinite sequence of higher charges that exist in this case.

\section*{Acknowledgments}

We would like to thank J. Avan, S. Das, C. I. Tan and A. Zhiboedov for interesting discussions. This work is supported by the Department of Energy under contract DE-FG02-91ER40688.

\end{document}

%% file: econfmacros.tex
\def\beq{\begin{equation}}
\def\eeq#1{\label{#1}\end{equation}}
\def\eeqn{\end{equation}}

\def\beqa{\begin{eqnarray}}
\def\eeqa#1{\label{#1}\end{eqnarray}}
\def\eeqan{\end{eqnarray}}

\let\bar=\overbar

\def\Dslash{\not{\hbox{\kern-4pt $D$}}}
\def\dslash{\not{\hbox{\kern-2pt $\del$}}}

\def\msb{{\bar{\ssstyle M \kern -1pt S}}}

